# Changes in the fine structure of stochastic distributions as a consequence of space-time fluctuations

Simon E. Shnoll

*Physical Department, Moscow State University, Moscow 119992, Russia Institute of Theoretical and Experimental Biophysics, Russian Academy of Sciences, Pushchino, Moscow Region, 142290, Russia*

Email: shnoll@iteb.ru

## Introduction.

Earlier we showed that the fine structure of the spectrum of amplitude variations in the results of measurements of the processes of different nature (in other words, the fine structure of the dispersion of results or the pattern of the corresponding histograms) is subject to "*macroscopic fluctuations*", changing regularly with time. These changes indicate that the "dispersion of results" that remains after all artifacts are excluded inevitably accompanies any measurements and reflects very basic features of our world. In our research, we have come to the conclusion that this dispersion of results is the effect of space-time fluctuations, which, in their turn, are caused by the movement of the measured object in an anisotropic gravitational field. Among other things, this conclusion means that the examination of the detailed pattern of distributions obtained from the results of measurement of the dynamics of processes of different nature discovers laws, which cannot be revealed with traditional methods for the analysis of time series.

These assertions are based on the results of long-term experimental investigations conducted for many decades. The major part of these results, starting with 1958, is published in Russian. The goal of this paper is to give a brief review of those results and provide corresponding references.

The most general conclusion of our research is the evidence that the fine structure of stochastic distributions is not accidental. In other words, noncasual is the pattern of histograms plotted from a rather small number of the results of measurement of the dynamics of processes of different nature, from the biochemical reactions and noise in the gravitational antenna to the α-decay [1-24].

## 1. The "effect of near zone".

The first evidence of the histogram pattern changing regularly in time is the "effect of near zone".

This effect means that similar histograms are significantly more probable to appear in the nearby (neighboring) intervals of the time series of the results of measurements. The similarity of the pattern of histograms plotted from independent intervals of a time series implies the presence of an external (towards the process studied) factor, which determines the pattern of the histogram. The independence of the "near zone" effect of the nature of the process indicates that this factor has a quite general nature.

## 2. Measurements of processes of different nature.

The second evidence comes from the similarity of the pattern of histograms plotted from the results of simultaneous independent measurements of processes of different nature at the same geographical point. In view of the fundamental difference in the nature of those processes and methods of their measuring, such a similarity also means that the factor, determining the histogram pattern, has a quite general nature. The similarity of histograms when under study are the processes, in which the ranges of transduced energy differ by dozens of orders (40 orders if the matter concerns the noise in the gravitational antenna and the alpha-decay), implies that this factor has no relation to energy.

## 3. Regular changes in the histogram patterns.

The third evidence of noncasuality of the histogram patterns is their regular changing with time. The regularities are revealed in the existence of the following periods in the change of the probability of similar histograms to appear.
3.1. Near-daily periods; these are well-resolvable "sidereal" (1436 min) and "solar" (1440 min) daily periods. These periods mean the dependence of the histogram pattern on the rotation of the Earth around its axis. The pattern is determined by two independent factors: the disposition relative to the starry sky and that relative to the Sun.
3.2. Approximately 27-day periods. These periods can be considered as an indication of the dependence of the histogram pattern on the disposition relative to the nearby celestial bodies: the Sun, the Moon and, probably, the planets.
3.3. Yearly periods; these are well-resolvable "calendar" (365 solar days) and "sidereal" (365 solar days plus 6 h and 9 min) yearly periods.
All these periods imply the dependence of the histogram pattern on the (1) rotation of the Earth around its axis and (2) movement of the Earth along its circumsolar orbit.

## 4. The observed local-time synchronism.

The dependence of the histogram pattern on the Earth rotation around its axis is clearly revealed in a phenomenon of "*synchronization at the local time*", when similar histograms are highly probable to appear at different geographical points (from Arctic to Antarctic, in the Western and Eastern hemispheres) at the same *local* time. It is astonishing that the local-time synchronism with the precision of 1 min is observed independently of the regional latitude at the most extreme distances – as extreme as possible on the Earth (about 15,000 km).

## 5. The synchronism observed at different latitudes.

The dependence of the histogram pattern on the Earth rotation around its axis is also revealed in the disappearance of the near-daily periods close to the North Pole. Such measurements were conducted at the latitude of 82 degree North in 2000. The analysis of histograms from the 15-min and 60-min segments showed no near-daily periods, but these periods keep in the sets of histograms plotted from the 1-min segments. Also keeping was the local-time synchronism in the appearance of similar histograms.

As follows from these results, it would be highly interesting to conduct measurements as close as possible to the North Pole. That was unfeasible, and we performed measurements with collimators, which channel α-particles, outgoing from a $^{239}$Pu sample, in a certain direction. The results of those experiments made us change our views fundamentally.

## 6. The collimator directed at the Pole Star.

Measurements were taken with the collimator directed at the Pole Star. In the analysis of histograms plotted from the results of counting α-particles that were traveling North (in the direction of the Pole Star), the near-daily periods were not observed, nor was the near-zone effect. The measurements were made in Pushchino (54. latitude North), but the effect is as would be expected at 90. North, i. e. at the North Pole. This means that the histogram pattern depends on the spatial direction of the process measured. Such dependence, in its turn, implies a sharp anisotropy of space. Additionally, it becomes clear that the matter does not concern any "effect" or "influence" on the object under examination. The case in point is changes, fluctuations of the space-time emerging from the rotation of the Earth around its axis and the movement of the planet along its circumsolar orbit [9, 13, 14, 15, 19, 20, 21].

## 7. The East and West-directed collimators.

This effect was confirmed in the experiments with two collimators, directed East and West correspondingly. In those experiments, two important effects were discovered.
7.1. The histograms registered in the experiments with the East-directed collimator ("east histograms") are similar to those "west histograms" that are delayed by 718 min, i.e. by a half of the sidereal day.
7.2. No similar histograms were observed at the simultaneous measurements with the "east" and "west" collimators. Without collimators, similar histograms are highly probable to appear at the same place and time. This space-time synchronism dissappears when counted are α-particles streaming in the different directions.

These results are in agreement with the concept that the histogram pattern depends on the vector of the alpha-particle outgo relative to a certain point at the coelosphere [20].

## 8. The experiments with the rotating collimators.

These investigations were naturally followed by the experiments with the rotating collimators [22, 24].
8.1. The collimator is rotating counter-clockwise (i.e., together with the Earth), the coelosphere is scanned with a period equal to the number of the collimator rotations per day plus one rotation made by the Earth itself. We examined the dependence of the probability of similar histograms to appear on the number of collimator rotations per day. Just as expected, the probability turned out to jump with periods equal to 1440 min divided by the number of collimator rotations per day plus 1. We evaluated data at 1, 2, 3, 4, 5, 6, 7, 11 and 23 collimator rotations per day and found periods equal to 12, 8, 6 etc. hours. The analysis of highly resolved data (with a resolution of 1 min) revealed that each of these periods had two extrema: "sidereal" and "solar". These results indicate that the histogram pattern is indeed determined by how the direction of the alpha-particle outgo relates to the "picture of the heaven" [24].

8.2. When the collimator made 1 clockwise rotation per day, the rotation of the Earth was compensated for (alpha-particles always outgo in the direction of the same region of the coelosphere) and, correspondingly, the daily periods disappeared. This result was completely analogous to the results of measurements near the North Pole and measurements with the immobile collimator directed at the Polar Star [20].

8.3. With the collimator placed in the ecliptic plane, directed at the Sun and making 1 clockwise rotation per day, alpha-particles will constantly outgo in the direction of the Sun. As was expected, the near-daily periods, both solar and sidereal, disappeared under such conditions.

## 9. The 718-min period.

The pattern of histograms is determined by a complex set of cosmo-physical factors. As follows from the existence of the near-27-day periods, among these factors may be the relative positions and states of the Sun, the Moon and the Earth. We repeatedly observed similar histograms during the rises and sets of the Sun and the Moon. A very large volume of work has been carried out. Yet, we have not found a histogram pattern, which would be characteristic for those moments. A review and analysis of the corresponding results will be given in a special paper. Here, I shall note one quite paradoxical result: on the equinoctial days, one can see a clear period in the appearance of similar histograms, which is equal to 718 min (i.e. a half of the sidereal day). There is no such a period on the solstice days. This phenomenon indicates that the histogram pattern depends on the ecliptic position of the Sun. If it is so, we can expect that on the equator, the period of 718 min will be observed year-round.

## 10. The observations during eclipses.

All the results presented above were obtained by the evaluation of tens of thousands of histogram pairs in every experiment. So, these results have a stochastic character. An absolutely different approach is used in the search for characteristic histogram patterns in the periods of the new moon and solar eclipses. In these cases, we go right to the analysis of the histogram patterns at a certain predetermined moment. Doing so, we have discovered an amazing phenomenon. At the moment of the new moon, a certain characteristic histogram appears practically simultaneously at different longitudes and latitudes – all over the Earth. This characteristic histogram corresponds to a time segment of 0.5-1.0 min [21]. When the solar eclipse is in the maximum (as a rule, this moment does not coincide with the time of the new moon), a specific histogram also appears; however, it has a different pattern. Such specific patterns emerge not only in the moments of the new moon or solar eclipses. But the probability of their appearance at these very moments at different places and on different dates (months, years) being accidental is extremely low. These specific patterns do not relate to the tidal effects. Nor they depend on the place on the Earth surface, where the moon shadow falls during the eclipse or the new moon.

# 11. The possible nature of "macroscopic fluctuations".

Above I have presented a brief review of the main phenomena that are united by the notion of "macroscopic fluctuations". A number of works suggested different hypotheses on the nature of those phenomena [3, 9, 10, 13-15, 19, 27-31], concerning some general categories such as discreteness and continuity, symmetry, the nature of numbers, stochasticity. In this section of the paper I would draw attention to the question of how some of the discovered phenomena can be considered in relation to these general categories.

11.1. Non-energetic nature of the phenomena. Space-time fluctuations [14, 19].
It is clear that we deal with non-energetic phenomena. As mentioned above, the ranges of energies in the biochemical reactions, noise in the gravitational antenna and α-decay differ by many orders. At the same time, the corresponding histogram patterns are similar with a high probability at the same local time at different geographical points. The only thing common for such different processes is the space-time, in which they occur. Therefore, the characteristics of the space-time change every next moment.

It is important to note that the "macroscopic fluctuations" do not result from the effect of any factors on the object under examination. They just reflect the state of the space-time.

The changes of the space-time can follow the alterations of the gravitational field. These alterations are determined by the movement of the examined object in a heterogeneous gravitational field. The heterogeneity results from the existence of "mass thicknesses", i.e. heavenly bodies. The movement includes the daily rotation of the Earth, its translocation along the circumsolar orbit and, probably, the drift of the solar system in the galaxy. All these forms of movement seem to be reflected in the correspondent periods of variation of histogram patterns. It is unclear how do the fluctuations of the space-time transform into the pattern of histograms.

11.2. Fractality [14, 19].
We suppose that the histogram pattern varies due to the change of the cosmo-physical conditions in the process of the Earth movement around its axis and along the circumsolar orbit. Then we might expect that the shorter are the intervals, for which histograms are plotted, the more similar would be the histogram patterns. This corresponds to the concept of "lifetime" of a certain idea of form. This concept is an obvious consequence of the "effect of near zone", when the probability of histogram patterns to be similar is higher for the histograms from the neighbor intervals.

However, we failed to find such a short interval that the histogram pattern "would not have time to change". The maximal probability of histograms to be similar only in the first, the nearest interval does not change upon variation of this interval from several hours to milliseconds. This phenomenon corresponds to the notion of "fractality"; however, the physical meaning of this fractality needs to be clarified.

As follows from the dependence of the histogram pattern on direction obtained in the experiments with collimators, we deal with a spatial heterogeneity on the scale of the order of $10^{-13}$ cm: the dependence of the histogram pattern should be determined before the outgo of alpha-particles from the nucleus. So to "stop the instant", stop the histogram changing we should have worked with the correspondingly small time intervals. Maybe, it will be possible someday.

### 11.3. The mirror symmetry, chirality of histograms [7].

Quite often (up to 30% of cases), the patterns of the successive histograms are reflection symmetric. There are right and left forms, and they may be very complex. This phenomenon possibly means that chirality is an immanent feature of the space-time.

### 11.4. "Stochasticity along abscissa and regularity along ordinate".

Our main results – the evidence of non-stochasticity of the fine structure of sampling distributions, i.e. the fine structure of the spectrum of amplitude fluctuations in the processes of any nature, i.e. the fine structure of the corresponding histograms – implies the existence of a particular class of macroscopic stochastic processes.

Among such processes is the radioactive decay. This is an "*a priori* stochastic" (i.e. stochastic according to the accepted criteria) process. However, the pattern of histograms (i.e. the fine structure of the amplitudes of fluctuations of the decay rate) changes regularly with time.

The point is that in the majority of cases, stochasticity is treated as an irregular succession of events. Succession in time, just one after another. This is "stochasticity along the axis of abscises".

For macroscopic processes, the distributions of the amplitudes of fluctuations of measured quantities are considered to correspond smooth distributions of Gauss-Poisson type. The available fitting criteria are integral, they are based on averaging, smoothing of those fluctuations. Such fitting criteria cannot "sense" the fine structure of distributions. According to these criteria, the processes we study, such as radioactive decay, well correspond to traditional views.

However, known for more then a hundred years is a noticeable exception – atomic spectra. While the transitions of electrons from one level to another are "*a priori* stochastic", the energies of the levels are sharply discrete. The "stochastic along the abscissa" process of transition is "regular along the ordinate".

The result of our work is the discovery of analogous macroscopic processes. In the process of fluctuating, the measured quantities take values, some of which are observed more often than the others; there are "forbidden" and "allowed" values of the measured quantities. This is what we see in the fine structure of histograms, with all its "peaks and troughs". The "macroscopic quantization" differs from the quantization in microworld. Here only the "idea of histogram form" remains invariant, whereas the concrete values, corresponding to extrema, can change. This is the main difference between the spectra of amplitude fluctuations of macroscopic processes and the atomic spectra.

### 11.5. The fine structure of histograms.

The presence of "peaks and troughs" in histogram patterns is a consequence of two causes: arithmetic (algorithmic) and physical [7, 14, 19].

### 11.5.1. The arithmetic or algorithmic cause of discreteness [7, 14, 19].

This phenomena lies in a very unequal number of factors (divisors) corresponding to the natural sequence. If the measured value is a result of operations based on the algorithms of division, multiplication, exponentiation, then discreteness will be unavoidable. Correspondingly, the histogram patterns will be determined by these algorithms. This can be seen, for example, in the computer simulation of the process of radioactive decay (Poisson statistics). The pattern of some histograms obtained in sich a simulation is indistinguishable from the pattern of histograms plotted for the radioactive decay data. However, the sequence of "computer" histogram patterns, in contrast to that of "physical" ones, does not depend on time and can be reproduced over and over again by launching the simulation program with the same parameters. This sequence is determined by the nature of numbers and the algorithms used. In our practice, we had an unusual incident when the sequence of histogram patterns created by a random number generator was similar, with high probability, to the sequence obtained from the radioactive decay data. If studied systematically, this case might give a clue to the nature of those "physical algorithms" that determine the time changes of the patterns of physical histograms [19].

### 11.5.2. The physical cause of discreteness is the interference of wave fluxes [19].
The fine structure of histograms, the presence of narrow extrema cannot have a probabilistic nature. According to Poisson statistics, which radioactive decay roughly obeys to, the width of such extrema should be of order $N^{1/2}$.

Therefore, if extrema that are neighbour in the histogram pattern have close values of N, they should overlap, but they do not. Such narrow extrema can arise only as a result of interference. Hence, the fine structure of histograms plotted from the results of measurements of any nature would be an interference of some waves. As follows from all the material presented above, the matter concerns processes caused by the movement of the Earth (and objects on its surface) relative to the "mass thicknesses". So it would be logical to define the waves whose interference is reflected in the histogram patterns as "gravitational".

The results of experiments with collimators, producing narrow beams of alpha-particles, lead us to the conclusion about the sharp anisotropy of our world. The corresponding wave fluxes should be very narrow.

Virtually, collimators are not necessary to reveal this anisotropy. We observe highly resolved daily and yearly periods in the changing of the probability of a certain histogram pattern to appear repeatedly (the resolution is 1 min). The histogram patterns specific for the new moon and solar eclipses can appear at different geographical points synchronously, with the accuracy of 0.5 min. The local-time synchronism at different geographical points (almost 15,000 km distant from each other) is also determined by a sharp extremum on the curve of distribution over intervals with the resolution of 1 min. In the experiments with the rotation of collimators, the "sidereal" and "solar" periods are also observed with the one-minute resolution.

Taken together, all these facts can mean that we deal with narrowly directed wave fluxes, "beams". The narrowness of these putative fluxes or beams exceeds the aperture of collimators. Collimators with the diameter of 0.9 mm and length of 10 mm isolate in the coelosphere a window of about 5°, this corresponding to approximately 20 min at the Earth's daily rotation rate. This fact noted by D.P. Kharakoz could be explained if we admit that the "beams" are more narrow than the aperture of our collimators.

Even with the fact that the matter concerns the changes of the histogram pattern and the movement of the Earth relative to the sphere of immobile stars, the Moon and the Sun, it is not necessary to consider anisotropy as being only due to the heterogeneous distribution of masses (presence of celestial bodies) in the space. It is possible that this anisotropy is caused by a preferential direction, which, for example, is due to the drift of the solar system to the constellation of Hercules. The existence of such a direction is an old problem of physics. In this connection, of great value for us are the results of interference experiments of Dayton Miller [43], the experiments and conception of Allais [42], de Witte's measurements [47] and R. Cahill's conception [44, 46]. It is necessary to mention that several years ago, V.K. Lyapidevsky [29] and I.M. Dmitrievsky [30] considered the preferential direction in space as the cause of the effects we observed.

In this case, we can say that for many years, we have studied phenomena indicating the existence of gravitational waves. Then the problem of registration of gravitational waves can be approached differently: instead of using bulky and expensive devices, such as Weber's antennas, one could register the changes of the fine structure of histograms plotted from the results of measurements of certain choosen processes. In this situation, we suppose that of principal importance are works of L.B. Borisova [50] and D.D. Rabunsky [49] on the theory and methods of registration of gravitational waves and the concept of "global scaling" advanced by Hartmut Muller [51].

## Acknowledgements.

I thank my colleagues I.A. Rubinshtein, V.A. Shlekhtarev, V.A. Kolombet, N.V. Udaltsova, E.V. Pozharsky, T.A. Zenchenko, K.I. Zenchenko, A.A. Konradov, L.M. Ovchinnikova, T.S. Malova, T.Ya. Britsina and N.P. Ivanova for many years of collaboration. I appreciate valuable comments and psychological support from D.P. Kharakoz, V.I. Bruskov, F.I. Ataullakhanov, V.N. Morozov, I.I. Berulis, B.M. Vladimirsky, V.K. Lyapidevsky, I.M. Dmitrievsky, B.U. Rodionov, S.N. Shapovalov, O.A. Troshichev, E.S. Gorshkov, A.V. Makarevich, V.A. Sadovnichy, U.S. Vladimirov, V.A. Namiot, N.G. Esipova, G.V. Lisichkin, Yu.A. Baurov, D.S. Chernavsky, B.V. Komberg, V.L. Ginzburg, E.L. Feinberg, G.T. Zatsepin. Invaluable advice, psychological and financial support of M.N. Kondrashova and V.P. Tikhonov are in the basis of the results obtained. I am grateful to G.M. Frank, G.R. Ivanitsky, E.E. Fesenko, V.A. Tverdislov for many years of patience and support. Special thanks to my respected teacher Sergey E. Severin and my older friend Lev A. Blumenfeld for the years of joyful conversations.

## References.